\documentclass[sigconf, nonacm, natbib]{acmart}

\usepackage{makecell}  
\usepackage{afterpage}
\usepackage{float}
\usepackage{multicol}
\usepackage{cuted}
\usepackage[skip=1pt]{caption}
\usepackage[inline]{enumitem}
\usepackage{hyperref}
\AtBeginDocument{%
  }

\makeatletter
\newcommand\headingnodot{\def\@toclevel{4}%
  \@startsection{paragraph}{4}{\z@}%
  {-.2\baselineskip \@plus -2\p@ \@minus -.2\p@}%
  {-3.5\p@}%
  {\ACM@NRadjust{\bfseries}}}

\makeatother

\begin{document}

\title{Contrastive Learning for Diversity-Aware Product Recommendations in Retail}

\author{Vasileios Karlis}
\authornote{The research was performed during the author’s internship at IKEA Retail under co-supervision of the University of Amsterdam. The author is currently employed by Elsevier B.V., Amsterdam, The Netherlands.}
\orcid{0009-0005-6068-180X}
\affiliation{%
  \institution{University of Amsterdam}
  \city{Amsterdam}
  \country{The Netherlands}}
\email{vasilis.karlis@student.uva.nl}

\author{Ezgi Yıldırım}
\authornote{Corresponding author.}
\authornote{\textcopyright Ingka Holding B.V. 2026. The IKEA logo and the IKEA wordmark are registered trademarks of Inter IKEA Systems B.V.}
\orcid{0000-0001-9684-5769}
\affiliation{%
  \institution{IKEA Retail (Ingka Group)}
  \city{Amsterdam}
  \country{The Netherlands}}
\email{ezgi.yildirim@ingka.ikea.com}

\author{David Vos}
\orcid{0009-0003-8925-1585}
\affiliation{%
  \institution{University of Amsterdam}
  \city{Amsterdam}
  \country{The Netherlands}}
  \email{d.j.a.vos@uva.nl}

\author{Maarten de Rijke}
\orcid{0000-0002-1086-0202}
 \affiliation{%
   \institution{University of Amsterdam}
   \city{Amsterdam}
   \country{The Netherlands}}
   \email{m.derijke@uva.nl}
\renewcommand{\shortauthors}{Karlis et al.}

\begin{abstract}
Recommender systems often struggle with long-tail distributions and limited item catalog exposure, where a small subset of popular items dominates recommendations. This challenge is especially critical in large-scale online retail settings with extensive and diverse product assortments. This paper introduces an approach to enhance catalog coverage without compromising recommendation quality in the existing digital recommendation pipeline at IKEA Retail.
Drawing inspiration from recent advances in negative sampling to address popularity bias, we integrate contrastive learning with carefully selected negative samples. Through offline and online evaluations, we demonstrate that our method improves catalog coverage, ensuring a more diverse set of recommendations yet preserving strong recommendation performance.
\end{abstract}

\maketitle
\section{Introduction}


Recommender systems have become a critical component for personalizing user experiences in e-commerce platforms \cite{10.1145/3370082}. An important aspect of these systems is \emph{catalog coverage}, which defines the ability to expose users to a broad range of products \cite{10.1145/1864708.1864761}. Without sufficient coverage, recommender systems risk reinforcing narrow, popularity-driven selections, limiting user discovery, and missing opportunities for both the platform and the user \cite{zhao2023fairness}.

\begin{figure}[h]
  \centering
  \includegraphics[width=0.70\linewidth]{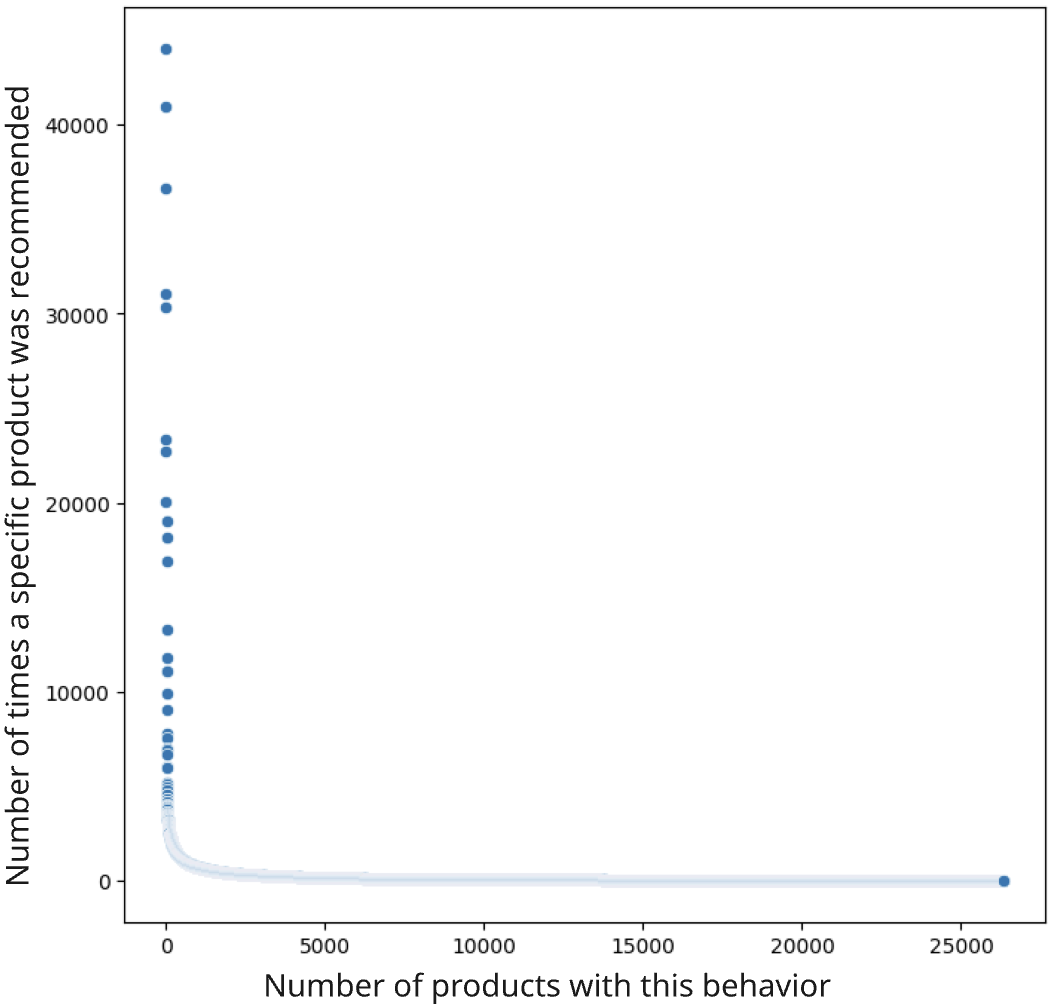}
  \caption{The long-tail of item popularity as captured in our recommendations for the Netherlands, in time period $t$.
   }
\label{fig:long-tail}
   \vspace*{-5mm}
\end{figure}

We focus on the recommender system currently deployed on \hyperlink{https://www.ikea.com/}{IKEA.com}\textcopyright. \autoref{fig:long-tail} shows the distribution of products across recommendations. As can be observed, recommendations of the current system show a prominent long-tail phenomenon. This skewed distribution implies that a small set of popular items dominate user interactions, while the majority of products (``long tail") receive limited attention. While retail demand typically follows a long-tail distribution, an ideal recommendation system balances revenue-driven head exposure with catalog diversity \cite{carlos2016netflix}. We aim for a softened power-law distribution, where high-performing items retain visibility, but tail items are surfaced to support discovery, personalization, and business diversity goals.

Our recommender system is an omni-channel and multi-session pipeline that serves personalized recommendations across devices and sessions. The system is structured around three primary components, as illustrated in \autoref{fig:product}:

\begin{figure*}
  \centering
  \includegraphics[width=\linewidth]{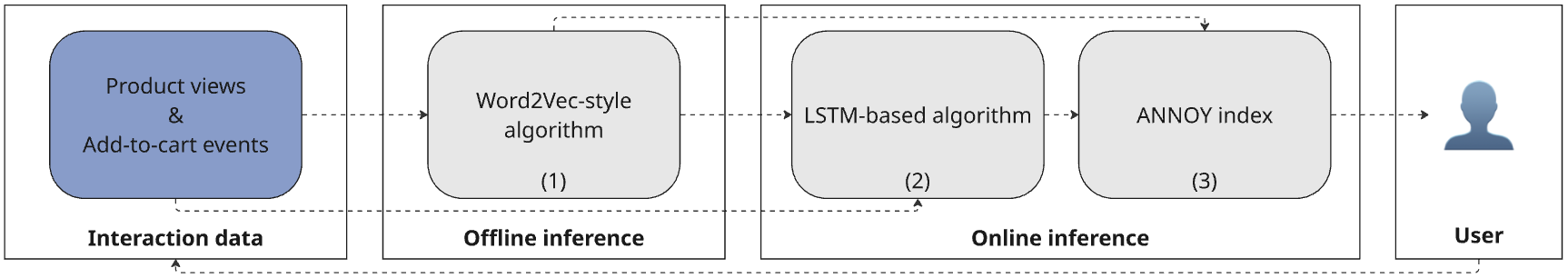}
  \caption{The main algorithmic components of the existing recommender system at IKEA Retail.}
   \label{fig:product}
\end{figure*}



\begin{enumerate}[leftmargin=*]
    \item \textbf{Product embedding generation.} 
    Embeddings for IKEA’s products are generated using a variant of a Word2Vec-style algorithm \cite{mikolov2013efficientestimationwordrepresentations}, which, instead of processing sequences of words, handles sequences of product ids from customer product views. These embeddings are computed offline and are static throughout the recommendation process, serving as the foundational representation of each product.
    

    \item \textbf{Session-based interaction modeling.} 
    To capture omni-chan\-nel and multi-session dynamics, an LSTM model is employed. This model is trained on sequences of product views as inputs and add-to-cart events as labels.  Product embeddings are used as a fixed input layer during this phase. The last hidden state of the LSTM is an approximate representation of the product that the user is most likely to add to their cart. The LSTM parameters are updated using a cosine similarity loss, quantifying the difference between the predicted embedding and the embedding of the product that was added to the cart.

    \item \textbf{Nearest neighbor search.} 
    The final component of the model uses a nearest neighbor search algorithm, ANNOY \cite{li2019approximate}, to identify and recommend products closest to the predicted product embedding from the previous step. This search procedure efficiently retrieves products that best align with the user’s preferences, ensuring relevant recommendations are made at every interaction. 
\end{enumerate}




\noindent%
In this work, we explore alternatives to the existing cosine similarity loss function to alleviate the long-tail distribution of IKEA’s recommendations, while preserving recommendation performance. Inspired by methods addressing popularity bias \cite{prakash2024evaluating,koneru2024enhancing}, we introduce contrastive learning with negative samples, extending the training objective beyond predicting embeddings close to the true label. We adapt existing contrastive learning methods to the pipeline of IKEA, aligning them with the existing cosine similarity loss function. Our negative sampling procedure encourages dissimilarity from strategically selected negative samples. 

Recent work on negative sampling in recommender systems \cite{kang2018selfattentivesequentialrecommendation} builds on architectures such as SASRec that optimize a probability distribution over the product catalog using a cross-entropy loss. Our setup optimizes the LSTM model with a cosine similarity loss, explicitly aligning predicted and ground-truth embeddings.

As our first contribution, we reformulate contrastive learning within our recommendation framework where the model predicts an embedding, instead of a probabilistic output.
Second, we evaluate two recent state-of-the-art negative sampling strategies in this new setting. We refer to these strategies as ``in-batch'' negative sampling, and a variation of adaptive sampling combined with ``top-$k$'' filtering, applied on top of the in-batch method \cite{prakash2024evaluating, koneru2024enhancing}. We achieve a notable improvement in catalog coverage, enhancing recommendation diversity without degrading ranking performance.


\section{Experimental Setup}
In this section, we first introduce modified loss functions to allow for contrastive learning with our cosine similarity loss. Second, we introduce different types of negative sampling. To evaluate the effectiveness of our approach\footnote{To ensure reproducibility of our experiments, we provide the following hyper-parameters selected in our setup based on their respective performance. For the weighted loss in \autoref{eq:weighted}, we set the hyper-parameters to $\alpha = 2$ and $\beta = 1$. For the cross-entropy loss in \autoref{eq:ce}, we use a temperature parameter of $\tau = 0.05$. For the top-$k$ negative sampling strategy, we apply a top-$k$ filter based on cosine similarity on the in-batch sample pool of $100$ negatives.}, we conduct experiments using both real-world data from IKEA’s online platform and a public benchmark dataset. We compare alternative loss functions and different negative sampling strategies, and lastly assess their impact using a combination of accuracy and beyond-accuracy metrics. 

\subsection{Loss functions}

Most recent work on sequential recommender systems literature applies variations of binary cross-entropy (BCE) for contrastive learning. However, in IKEA’s pipeline, we adopt a different setup. The model is trained by computing the cosine similarity between the last hidden state of an LSTM-based architecture and the ground-truth embedding. This motivates the use of cosine similarity as a core training objective when exploring contrastive learning within the existing setup. We propose two variants of the cosine similarity loss. Below, we describe each loss formulation, where $sim(x, y)$ denotes the cosine similarity between vectors $x$ and $y$.

\begin{itemize}[leftmargin=*]
\item \textbf{Cosine similarity loss.}
Our baseline loss uses classic cosine similarity without incorporating negative samples. The model learns to maximize the similarity between its output $z_i$ and the embedding of the true label $z_t$:
\begin{equation}
\label{eq:cosine}
L_i = - sim(z_i, z_t) 
\end{equation}

\item \textbf{Weighted loss.}
This loss function introduces contrastive learning by penalizing the model if its output $z_i$ is too similar to negative samples $z_j$. The strength of this penalty is controlled by hyperparameters $\alpha$ and $\beta$.
%
\begin{equation} 
\label{eq:weighted}
L_i = - \alpha \cdot sim(z_i, z_t) + \beta \cdot \frac{1}{N} \sum_{j=1}^{N} sim(z_j, z_f),
\end{equation}
where $N$ is the number of negative samples and $z_f$ is the negative label.

\item \textbf{Cross-entropy loss.}
Inspired by \cite{chen2020simple,gao2021simcse}, this loss function formulates contrastive learning as a classification task over in-batch negative samples. The model is trained to identify the true label, $z_t$, among $N$ candidates, with $\tau$ being a temperature parameter:
\begin{equation}
\label{eq:ce}
L_i = - \log{\frac{\exp(sim(z_i, z_t) / \tau)}{\sum_{j=1}^{N} \exp(sim(z_j, z_f) / \tau)}}.
\end{equation} 
\end{itemize}

\begin{table*}
  \caption{Experimental results for the IKEA and RetailRocket datasets.}
  \label{tbl:eval}
  \begin{tabular}{lll ccc ccc}

    \toprule
    \multicolumn{3}{c}{\textbf{Method}} & \multicolumn{3}{c}{\textbf{IKEA dataset}} & \multicolumn{3}{c}{\textbf{RetailRocket dataset}} \\
    \cmidrule(lr){1-3} \cmidrule(lr){4-6} \cmidrule(lr){7-9}
    Loss function & Sampling strategy & Sampling size & NDCG@10 & GC & Coverage & NDCG@10 & GC & Coverage \\
    \midrule
    Cosine similarity & NA & NA & 0.1125 & 0.8745 & 0.2722 & 0.0002 & 0.9988 & 0.0080 \\
    Weighted & In-batch & 100 & 0.1093 & \textbf{0.814} & 0.3040 & 0.0018 & 0.9908 & 0.0319 \\
    Weighted & In-batch & 5 & 0.1091 & 0.8484 & 0.3047 & 0.0016 & 0.9922 & 0.0303 \\
    Weighted & Top-$k$ & 5 & \textbf{0.1198} & 0.8397 & 0.3164 & \textbf{0.0025} & \textbf{0.9849} & \textbf{0.0476} \\
    Cross-entropy & In-batch & 100 & \textbf{0.1198} & 0.8170 & \textbf{0.3441} & 0.0007 & 0.9934 & 0.0275 \\
    Cross-entropy & In-batch & 5 & 0.0907 & 0.8493 & 0.3067 & 0.0000 & 0.9851 & 0.0458 \\
    Cross-entropy & Top-$k$ & 5 & 0.1141 & 0.8212 & 0.3392 & 0.0002 & 0.9969 & 0.0171 \\
    
    \bottomrule
  \end{tabular}
\end{table*}

\subsection{Negative sampling strategies}
To enable contrastive learning within our embedding-based recommendation setup, we explore two negative sampling techniques inspired by recent literature, adapting them to fit the specific structure and constraints of our pipeline.

\begin{itemize}[leftmargin=*]
\item \textbf{In-batch negative sampling with sampling cap.} 
As a starting point, we adopt a variation of the in-batch negative sampling strategy introduced by \citet{prakash2024evaluating}. In this approach, all other positive items within the current training batch are treated as negatives, excluding items from the user’s input sequence to avoid false negatives. Unlike the original formulation, where all batch items are considered, we introduce a configurable hyperparameter, a.k.a. ``sampling cap'', to limit the number of negatives used during training. This allows us to balance computational cost with learning signal strength.

\item \textbf{Adaptive top-$k$ in-batch negative sampling.}
Building on the previous technique, we investigate an adaptive sampling strategy, motivated by its performance in \cite{prakash2024evaluating,koneru2024enhancing}. In our implementation, we operate exclusively on in-batch negatives. This method dynamically selects the most informative negatives during training. We begin with in-batch negatives and compute their cosine similarity with the predicted embedding. The top-$k$ items, those most similar to the model’s output, are selected as hard negatives. Only these are used in the backward pass, enabling more focused updates and reducing noise from less informative samples. This selective strategy ensures that the model learns to distinguish between fine-grained differences in product embeddings. 
\end{itemize}


\subsection{Datasets}
We conduct our experiments using two datasets: (i)~a proprietary dataset from IKEA Netherlands and (ii)~the publicly available RetailRocket dataset~\cite{prakash2024evaluating}. In both cases, we use add-to-cart events as labels and represent the input as a sequence of previously viewed items, truncated to a maximum length of $100$.

To apply a consistent approach across both datasets, we generate product representations for the RetailRocket dataset using the same Word2Vec-style algorithm employed for the IKEA data.

For consistency in evaluation, we adopt the same data splitting strategy for both datasets. The data is first split chronologically into two parts: one for training the Word2Vec model, and the other for training and evaluating the LSTM-based recommendation model. The second part is further divided into training and evaluation subsets, ensuring that all user actions are kept within a single split to prevent data leakage.


\subsection{Evaluation metrics}
To evaluate our approach, we report both accuracy and beyond-accuracy metrics, in line with our goal of improving catalog coverage without compromising recommendation performance.

For accuracy-based evaluation, we use Normalized Discounted Cumulative Gain (NDCG) \cite{10.1145/3460231.3478848}, which measures the quality of the ranking by considering the position of the correct item in the recommended list. Since our task involves predicting a single add-to-cart item for each input sequence, NDCG@10 is well-suited to reflect the model’s ranking performance.

To assess diversity-oriented performance, we report two aggregate metrics: the Gini Coefficient (GC) and catalog coverage.
GC evaluates how equitably recommendations are distributed across the product catalog, capturing the extent of popularity bias. A value of $0$ indicates perfect equality in exposure, whereas $1$ denotes that a small subset of products dominates the recommendation output. We compute GC based on the distribution of recommended items in the test set, following~\cite{braun2023metrics}. We target a Gini coefficient in the range of $[0.4, 0.7]$, as recommended by \cite{do2022optimizing}, to achieve a desirable trade-off between exposure fairness and relevance. This range reflects a softened power-law distribution, maintaining visibility for high-performing products while enabling the discovery and personalization of tail items. Catalog coverage measures the proportion of unique items from the catalog that appear in at least one recommendation across the test set \cite{10.1145/1864708.1864761}. This metric reflects the system’s ability to expose users to a wider range of products. Coverage values range from $0$ (low exposure) to $1$ (high exposure).

\section{Results}
Exploring the results presented in \autoref{tbl:eval}, we observe that incorporating negative sampling into the training procedure consistently improves diversity-oriented metrics, such as coverage and Gini Coefficient, while also enhancing accuracy-based metrics like NDCG. 


For the IKEA dataset, we see that the best-performing method (cross-entropy loss with 100 in-batch negative samples) results in a 26.4\% improvement in catalog coverage, a 6.5\% improvement in Gini Coefficient, and a 6.4\% improvement in NDCG@10.We attribute the low NDCG@10 scores to the high cardinality of our recommendation setting, where only a single positive item is labeled among a large number of candidates, as noted by \cite{gruson2019offline} in the context of playlist recommendation algorithms. In an online evaluation performed on the IKEA Netherlands website by using the weighted loss with 100 in-batch negatives, we observed a 2.53\% improvement in catalog coverage and a 9.59\% improvement in diversity over the baseline, which supports the offline results.
Similarly, for the RetailRocket dataset, the best-performing method (weighted loss with top-5 in-batch negative samples) significantly improves all metrics. 
Additionally, we find that our best-performing method yields an NGCG@10 value of 0.0025, which is substantially lower than the NDCG@10 value reported in \cite{prakash2024evaluating} (0.0302). This discrepancy is likely due to differences in our pipeline architecture and the data reduction introduced during the product embedding generation step. The best-performing methods vary between the two datasets, highlighting the importance of selecting the appropriate loss function, negative sampling strategy, and the number of negatives. 


    

\section{Conclusion}
We have found that in-batch negative sampling and a variation of adaptive sampling combined
with top-$k$ filtering, applied on top of the in-batch method, leads to notable improvement in catalog coverage, enhancing recommendation diversity without degrading ranking performance in our recommender system.

In future work, we plan to focus on negative sampling strategies to help the model learn user preferences more efficiently. Rather than selecting the most difficult examples based on recommendation performance, we aim to identify the most informative negative samples that contribute more to the learning curve for computational efficiency.

\vspace{0.5cm}

\begin{acks}
    This research was (partially) supported by the Dutch Research Council (NWO), under project numbers 024.004.022, NWA.1389.20.\-183, and KICH3.LTP.20.006, and the European Union under grant agreements No. 101070212 (FINDHR) and No. 101201510 (UNITE).
    Views and opinions expressed are those of the author(s) only and do not necessarily reflect those of their respective employers, funders and/or granting authorities.

\end{acks}

\bibliographystyle{ACM-Reference-Format}
\balance
\bibliography{references}


\end{document}